\documentclass[12pt]{iopart}
\usepackage{graphicx}
\usepackage{gensymb}
\usepackage{multirow}
\usepackage{float}

\begin{document}
\title{Sub-leading power corrections to transition form factors of $\eta$ and $\eta^{\prime}$ mesons from twist-six contributions}
\author{Janardan P Singh$^1$\footnote{Retired from services of The Maharaja Sayajirao University of Baroda}, Shesha D Patel$^2$}
\address{$^1$ $^2$Physics Department, Faculty of Science, The Maharaja Sayajirao University of Baroda, Vadodara 390002, Gujarat, INDIA}
\ead{$^1$janardanmsu@yahoo.com $^2$sheshapatel30@gmail.com}
\vspace{10pt}
\begin{indented}
\item[16] May 2018
\end{indented}

\begin{abstract}
Using collinear factorization, we have calculated twist-six corrections to $\eta\gamma$ and $\eta^{\prime}\gamma$ transition form factors in QCD. Only a sub-set of light-cone operators of twist-six which can be factorized as a product of two gauge invariant operators has been considered. Resulting power corrections start as $1/Q^{4}$  but coefficients are smaller compared to their counterparts for twist-four operators. Matrix element of pseudoscalar quark densities for $u$- and $d$-quarks between vacuum and meson state, $h_{q}$, has considerable influence on the result. Results obtained by superimposing our results on those obtained in literature for leading and sub-leading power corrections for lower twists have been plotted and compared with data.
\end{abstract}
\vspace{2pc}

\submitto{\jpg}
\maketitle

\section{Introduction}
Transition form factors (TFFs) of light pseudoscalar mesons have been extensively studied in last few decades experimentally as well as theoretically \cite{Behrend:1991,Gronberg:1998,Aubert:2006,Aubert:2009,Sanchez:2011,Uehara:2012,Lepage:1980,Efremov:1980,Braaten:1983,Ali:2003,Kroll:2003,Agaev:2004,Chernyak:1984,Agaev:2010,Agaev:2011,Kroll:2013,Agaev:2014,Khodjamirian:1999,Noguera:2012,Choi:2017}. These are known to be the simplest exclusive processes involving  strong interactions. They provide useful precision tests for the standard model, in particular, the QCD. A special role of TFFs of $\eta$ and $\eta^{\prime}$ is to determine their quark-gluon structure and to fix their distribution amplitudes (DAs). $\eta$ and $\eta^{\prime}$ mesons are characteristically different from pion in some sense. Due to SU(3) breaking, $\eta$ and $\eta^{\prime}$ states are mixtures of flavor-octet and flavor-singlet states, usually described by a mixing matrix \cite{Feldmann:1999,Charng:2006,Ball:2007,Singh:2013,Singh:2009}. Moreover, flavor singlet states can mix with two gluon states under evolution producing gluonic admixtures to both $\eta$ and $\eta^{\prime}$ mesons \cite{Ali:2003,Kroll:2003}. Axial U(1) anomaly gives rise to large mass to $\eta^{\prime}$ meson \cite{Witten:1979}.
\par TFFs enter predictions of several observables, such as the rates of rare decays $M\rightarrow l\overline{l}, (l=e,\mu)$ and hadronic light-by-light scattering contribution to the muon anomalous magnetic moment. Quark and gluon components of $\eta$ and $\eta^{\prime}$ meson DAs are necessary input information in theoretical calculations of various exclusive processes, such as computation of $B\rightarrow\eta^{(\prime)}$ TFFs \cite{Charng:2006,Ball:2007}, $\chi_{cJ}$ decays in pairs of $\eta$ and $\eta^{\prime}$ mesons, etc.
\par The standard approach for calculation of electromagnetic TFFs at high energy is based on collinear factorization \cite{Lepage:1980,Efremov:1980}: TFFs are convolution of perturbative hard-scattering amplitude and universal meson DAs which incorporate non-perturbative dynamics of QCD bound states. Meson DAs are either extracted from fits to data, or evaluated basing on low-energy models of the meson. DAs depend on the functions of the longitudinal momentum of the meson carried by partons inside the meson, power counting is achieved using a twist expansion of the operators on the light-cone. Since most of the experimental information is available for an asymmetric configuration in which one of the photon is quasi-on-mass-shell having a hadronic content, this requires non-perturbative input. In another approach \cite{Kurimoto:2001}, TFFs for $\pi^{0}$ have been calculated using transverse momentum dependent meson wave function or $k_{T}$-factorization. There is no direct interpretation of these results with those obtained using DAs. In the third approach \cite{Klopot:2011}, the photon TFFs of pseudoscalar mesons have been studied by means of anomaly sum rules which are based on the dispersive representation of the axial anomaly; this method does not rely on factorization. In addition, there are approaches which are not directly based on QCD: non-local chiral quark model \cite{Dorokhov:2011}, light-front quark model \cite{Choi:2017}, light-front holographic QCD \cite{Brodsky:2011}, combined analysis of low and high $Q^2$ data \cite{Noguera}, etc.
\par In view of the ongoing debate on the question whether hard exclusive hadronic reactions are under theoretical control, it is necessary to have a better understanding of the sub-leading power terms in the large momentum expansion. This will also be relevant for future high-intensity, medium energy experiments. The first approach based on factorization of short and long distance dynamics, is more versatile and has been extensively used in literature. As is well known, in this approach the quantity of interest is expanded in terms of twists of the operators. The TFFs of the $\eta$-and $\eta^{\prime}$-mesons have been calculated up to twist-4 so far. In this work we extend this calculation up to twist-6 operators. Simultaneously, we also include meson mass and quark mass corrections which give rise to $SU(3)_{F}$ breaking corrections.
\par This paper is organised as follows. In Sec. 2, we have given a short review of $\eta$-$\eta{\prime}$ mixing and DAs which have been used in this work. In Sec. 3, details of our calculation for TFFs have been presented. In Sec. 4, we have given numerical analysis of the theoretical results obtained: The results have been plotted individually, compared with the corresponding results for twist-4 correction, and compared the results obtained by superimposing our results on those obtained in literature for lower twists with data. Finally, in Sec. 5, we have summarized the work and concluded. Some more details of the quantities used in the main text have been relegated to an Appendix.

\section{$\eta$ - $\eta^{\prime}$ mixing and distribution amplitudes}
There are two equivalent ways in which $\eta-\eta^{\prime}$ mixing has been described in the literature. The quark-flavor mixing is convenient for this work and will be dealt with first. The decay constants $f_{M}^{(i)}$ for a meson M= $\eta$-$\eta^{\prime}$ are defined as \cite{Feldmann:1999},
\begin{equation}
\fl \langle M(p)\mid\frac{1}{\sqrt{2}}({\overline{u}(0)\gamma _{\mu}\gamma_{5}u(0)+\overline{d}(0)\gamma _{\mu}\gamma_{5}d(0)})\mid0\rangle
=-i f_{M}^{(q)}p_{\mu},
\end{equation}
\begin{equation}
\fl \langle M(p)\mid \overline{s}(0)\gamma _{\mu}\gamma_{5}s(0)\mid 0\rangle =-i f_{M}^{(s)}p_{\mu}.
\end{equation}
The leading-twist (twist-2) DAs are defined in terms of matrix elements of bilocal quark currents as,
\begin{equation}
\fl \langle M(p)\mid\frac{1}{\sqrt{2}}({\overline{u}(z_2)\gamma _{\mu}\gamma_{5}u(z_1)+\overline{d}(z_2)\gamma _{\mu}\gamma_{5}d(z_1)})\mid0\rangle = -i f_{M}^{(q)}p_{\mu}\int_0^{1}du e^{i(up.z_{2}+ \overline{u}p.z_{1})}\Phi_{M}^{(q)}(u)+....,
\end{equation}
\begin{equation}
\fl \langle M(p)\mid \overline{s}(z_2)\gamma _{\mu}\gamma_{5}s(z_1)\mid 0 \rangle = -i f_{M}^{(s)}p_{\mu}\int_0^{1}du e^{i(up.z_{2}+\overline{u}p.z_{1})}\Phi_{M}^{(s)}(u)+.... .
\end{equation}
In above equations as well as in the following, path-ordered gauge connection between non-local quark (gluon) operators \cite{Ball:1999,Ali:2003,Kroll:2003,Agaev:2004,Chernyak:1984,Agaev:2010,Agaev:2011,Kroll:2013,Agaev:2014,Khodjamirian:1999} is understood.
Left out terms on the R.H.S are either of higher twists or of higher order in light cone expansion.
The gluonic twist-2 DAs are defined as \cite{Kroll:2003},
\begin{equation}
\fl \langle M(p)\mid G_{\mu z}(z)\tilde{G}^{\mu z}(-z)|0\rangle=\frac{1}{2}(p.z)^{2}\frac{C_{F}}{\sqrt{3}}f_{M}^{0}\int_{0}^{1}du e^{ip.z(2u-1)}\Phi_{M}^{(g)}(u),
\end{equation}
where $z^{\mu}$ is a light-like vector. $G_{\mu z}$=$G_{\mu\xi}z^{\xi}$, $C_{F}=\frac{4}{3}$ and $f_{M}^{0}$ is singlet current decay constant. $\Phi_{M}^{(g)}(u)$ is antisymmetric under $u\leftrightarrow\overline{u}$.

The twist-2 DAs for quark-antiquark components and gluonic ones can be expanded in terms of Gegenbauer polynomials as \cite{Beneke:2003,Ali:2003,Kroll:2003}:
\begin{equation}
\fl \Phi_{M}^{i}(u,\mu)= 6u\overline{u}\left(1+\sum_{n=2,4,..}B_{n}^{M,i} (\mu)C_{n}^{3/2}(2u-1)\right), (i= q,s),
\end{equation}
\begin{equation}
\fl \Phi_{M}^{(g)}(u,\mu)=u^{2}\overline{u}^{2}\sum_{n=2,4,..}B_{n}^{M,g}(\mu)C_{n-1}^{5/2}(2u-1),
\end{equation}
where $\overline{u}$=$(1-u)$.  The quark-antiquark DAs are normalized as
\begin{equation}
\fl \int_{0}^{1} du \Phi_{M}^{(i)} (u)=1,
\end{equation}
whereas gluonic DA satisfies
\begin{equation}
\fl \int_{0}^{1} du \Phi_{M}^{(g)} (u)=0.
\end{equation}
Decay constants $f_M^{(i)}$ are usually parameterized as \cite{Feldmann:2000},
\begin{equation}\label{feta}
\fl \left(\begin{array}{cc} f_\eta^{(q)} & f_\eta^{(s)}\\
f_{\eta'}^{(q)} & f_{\eta'}^{(s)} \end{array}\right)
=\left(\begin{array}{cc} cos\phi & -sin\phi\\
sin\phi & cos\phi \end{array}\right)
\left(\begin{array}{cc} f_q & 0\\
0 & f_s \end{array}\right).
\end{equation}
Two-particle twist-3 DAs are introduced as follows \cite{Ball:1999,Ali:2003,Agaev:2014}:
\begin{equation}
\fl 2m_{q}\langle M(p)\mid\frac{1}{\sqrt{2}}\left(\overline{u}(z_{2})i \gamma_{5}u(z_{1})+\overline{d}(z_{2})i \gamma_{5}d(z_{1})\right)|0\rangle =\int_{0}^{1}du e^{i (up.z_{2}+\overline{u}p.z_{1})}\Phi_{3M}^{(q)p}(u),
\end{equation}
\begin{equation}
\fl 2m_{s}\langle M(p)\mid(\overline{s}(z_{2})i \gamma_{5}s(z_{1}))|0\rangle=
\int_{0}^{1}du e^{i (up.z_{2}+\overline{u}p.z_{1})}\Phi_{3M}^{(s)p}(u),
\end{equation}
\begin{eqnarray}
\fl 2m_{q}\langle M(p)\mid\frac{1}{\sqrt{2}}\left(\overline{u}(z_{2})\sigma_{\mu\nu}\gamma_{5}u(z_{1})+\overline{d}(z_{2})\sigma_{\mu\nu}\gamma_{5}d(z_{1})\right)|0\rangle=\frac{i}{6}(p_{\mu}z_{\nu}- p_{\nu}z_{\mu})\int_{0}^{1} du \times \nonumber\\ e^{i (up.z_{2}+\overline{u}p.z_{1})}\Phi_{3M}^{(q)\sigma}(u), \label{eq15}
\end{eqnarray}
\begin{equation}
\fl 2 m_{s}\langle M(p)\mid(\overline{s}(z_{2})\sigma_{\mu\nu}\gamma_{5}s(z_{1}))|0\rangle=\frac{i}{6}(p_{\mu}z_{\nu}- p_{\nu}z_{\mu})\int_{0}^{1} due^{i(up.z_{2}+\overline{u}p.z_{1})}\Phi_{3M}^{(s)\sigma}(u),
\end{equation}
where $ z=z_{2}-z_{1}$.
These twist-3 DAs are normalized as:
\numparts
\begin{eqnarray}
\fl \int_{0}^{1}du\Phi_{3M}^{(q)p}(u)&=&\int_{0}^{1}du\Phi_{3M}^{(q)\sigma} (u)=h_{M}^{(q)},
\\
\fl \int_{0}^{1}du\Phi_{3M}^{(s)p}(u)&=&\int_{0}^{1}du\Phi_{3M}^{(s)\sigma}(u)=h_{M}^{(s)}.
\end{eqnarray}
\endnumparts
\\We follow the following parametrization of these DAs \cite{Agaev:2014}:
\begin{equation}
\fl \left(
  \begin{array}{cc}
    \Phi_{3\eta}^{(q)p,\sigma} & \Phi_{3\eta}^{(s)p,\sigma} \\
    \Phi_{3\eta^{\prime}}^{(q)p,\sigma} &   \Phi_{3\eta^{\prime}}^{(s)p,\sigma}  \\
  \end{array}
\right)=U(\phi)\left(
          \begin{array}{cc}
            \Phi_{3q}^{p,\sigma} & 0 \\
            0 & \Phi_{3s}^{p,\sigma} \\
          \end{array}
        \right),
\end{equation}
where $U(\phi)$ is the same matrix as used in parameterizing $f_{M}^{(i)}$.
\\These DAs are expanded in terms of Gegenbauer polynomials as \cite{Agaev:2014}:
\numparts
\begin{eqnarray}
\fl \Phi_{3q}^{p}(u)&=&h_q+60 m_qf_{3q}C_{2}^{1/2}(2u-1)+...... ,
\\
\fl \Phi_{3s}^{p}(u)&=&h_s+60 m_sf_{3s}C_{2}^{1/2}(2u-1)+...... ,
\end{eqnarray}
\endnumparts
\numparts
\begin{eqnarray}
\fl \Phi_{3q}^{\sigma}(u)&=&6u\overline{u}\left(h_q+10m_qf_{3q}C_{2}^{3/2}(2u-1)+...\right),
\\
\fl \Phi_{3s}^{\sigma}(u)&=&6u\overline{u}\left(h_s+10m_sf_{3s}C_{2}^{3/2}(2u-1)+...\right).
\end{eqnarray}
\endnumparts
\\Constants are related as,
\begin{equation}\label{heta}
\fl \left(
  \begin{array}{cc}
    h_{\eta}^{(q)} & h_{\eta}^{(s)} \\
    h_{\eta^{\prime}}^{(q)} &  h_{\eta^{\prime}}^{(s)}  \\
  \end{array}
\right)=U(\phi)\left(
          \begin{array}{cc}
            h_{q} & 0 \\
            0 & h_{s} \\
          \end{array}
        \right),
\end{equation}
\begin{equation}
\fl \left(
  \begin{array}{cc}
    f_{3\eta}^{(q)} & f_{3\eta}^{(s)} \\
    f_{3\eta^{\prime}}^{(q)} &  f_{3\eta^{\prime}}^{(s)}  \\
  \end{array}
\right)=U(\phi)\left(
          \begin{array}{cc}
            f_{3q} & 0 \\
            0 & f_{3s} \\
          \end{array}
        \right).
\end{equation}
Combining Eqs. (\ref{feta}), (\ref{heta}) and (\ref{eq15}) one obtains
\begin{equation}\label{heta1}
\fl h_{q}=f_{q}\left( m_{\eta}^{2}\cos^{2}\phi+ m_{\eta^{\prime}}^{2} \sin^{2}\phi\right)-\sqrt{2}f_{s}\left(m_{\eta^{\prime}}^{2}-m_{\eta}^{2}\right)\sin\phi\cos\phi.
\end{equation}
Though $h_{q}$ itself is small, the combination in which it appears normally is $\frac{h_{q}}{m_{q}}$ which is not small. The constant $f_{{3q},{3s}}$ is three-particle decay constant which appears in twist-3 three particle DAs $\Phi_{3M}$ (see Appendix).
They are introduced in analogy with $f_{3\pi}$ as follows:
\begin{equation}
\fl \langle M(p)\mid\frac{1}{\sqrt{2}}({\overline{u}(0)\sigma _{z\mu}\gamma_{5}gG^{z\mu}u(0)+\overline{d}(0)\sigma _{z\mu}\gamma_{5}g G^{z\mu}d(0)})\mid0\rangle
=2i(p.z)^{2}f_{3M}^{(q)},
\end{equation}
\begin{equation}
\fl \langle M(p)\mid \overline{s}(0)\sigma _{z\mu}\gamma_{5} g G^{z\mu}s(0)\mid 0\rangle =2i(p.z)^{2}f_{3M}^{(s)}.
\end{equation}
The three distribution amplitudes of twist-3 $\Phi_{3M}^{p}$, $\Phi_{3M}^{\sigma}$ and $\Phi_{3M}$ are related to each other by equations of motion\cite{Braun:1990}.

\section{Twist-six contribution to $\gamma\gamma^*\rightarrow\eta ,\eta^{\prime}$ transition form factors}
\begin{figure*}
\includegraphics[width=\textwidth]{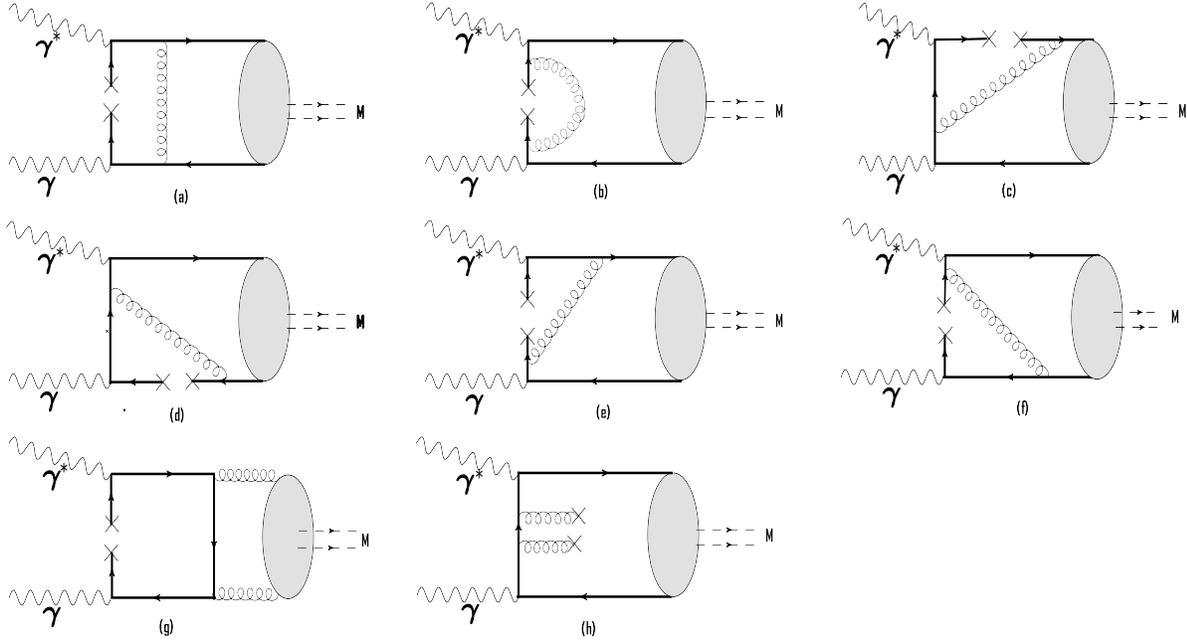}
\caption{Factorizable twist-six corrections to the TFF $F_{\gamma\gamma^{*}\rightarrow\eta^{(\prime)}}(Q^{2})$. The gluon condensate contribution (h) vanish.}
\end{figure*}
The meson transition form factor $F_{\gamma\gamma^{*}\rightarrow M}(q^2,(p-q)^2)$, $M$=$\eta, \eta^{\prime}$ is described by the following matrix element \cite{Agaev:2004,Chernyak:1984,Agaev:2010,Agaev:2011,Agaev:2014}.
\begin{equation}
\fl T_{\mu\nu}(q,p)\equiv i \int d^{4}x\exp^{-i qx}
 \langle M(p)\mid T\{j_{\mu}^{em}(x)j_{\nu}^{em}(0)\}\mid0\rangle  =e^2\epsilon_{\mu\nu\alpha\beta}q^{\alpha}p^{\beta}F_{\gamma^*\gamma^*\rightarrow M}(q^2,(p-q)^2),
\end{equation}
where
\begin{equation}
\fl j_{\mu}^{em}(x)=\sum_{\psi=u,d,s} e_{\psi}\overline{\psi}(x)\gamma_{\mu}\psi(x).
\end{equation}
We consider space-like form factors with $-q^2$=$Q^2$ being large and $(p-q)^{2}\approx0$. In such a situation we will denote the TFF $F_{\gamma^*\gamma^{*}\rightarrow M}(q^2,(p-q)^2)$ simply by $F_{\gamma\gamma^{*}\rightarrow M}(Q^2)$. The correlation function $T_{\mu\nu}(q,p)$ is dominated by light-like distances and therefore amendable to an expansion around the light-cone. The light-cone expansion is performed by integrating out the transverse and "minus" degrees of freedom and leaving the longitudinal momenta of the partons as the required degree of freedom. In practice, the transverse momenta are integrated only up to a cut-off, $\mu$, and momenta below $\mu$ are included in DAs. TFFs of pseudoscalar mesons have been calculated using factorization mainly upto twist-4 order. In Ref. \cite{Agaev:2011} an attempt has been made to extend the results for $\pi^{0}$ upto twist-6. We make a similar attempt here for $\eta$- and $\eta^{\prime}$- mesons. Since s-quark is involved in this case and the masses of $\eta$- and $\eta^{\prime}$- mesons are substantially larger compared to pion mass, we also introduce lowest order corrections arising due to finite s- quark mass and meson masses.
Several light-cone operators of twist-6 can be factorized as a product of two gauge-invariant twist-3 operators, or one twist-2 and one twist-4 operator; placed between vacuum and one meson state, such operators in factorization approximation can be evaluated as a product of quark or gluon condensate and twist-3 or twist-2 (along with quark mass) DA. Non-factorizable operators give rise to twist-6 multiparton meson DAs. It has been argued that due to conformal symmetry higher Fock states are strongly suppressed at $u\rightarrow 1$ giving negligible contributions\cite{Braun:2000}. We use this approach to estimate twist-6 contributions in this work.
The contribution to the matrix element $T_{\mu\nu}$ from the Feynman diagram given by Fig. 1(a) comes from bilocal pseudoscalar  operator and is given by
\begin{equation}\label{fig1a}
\fl T_{\mu\nu}^{(a)} (q,p)_{(pseudo)}=\frac{8 g^2}{9 q^2 m_{\rho}^2}\sum_{\psi=u,d,s}e_{\psi}^2\frac{1}{m_{\psi}}\langle\overline{\psi} \psi\rangle\epsilon_{\mu\nu\alpha\beta}q^\alpha p^\beta \int_{0}^{1} du \Phi_{3M}^{(\psi)p}(u)\frac{1}{(q-up)^2}.
\end{equation}
In the above equation we have used vector meson dominance model and replaced $(q-p)^2\rightarrow m_{\rho}^2$ in the denominator. We expand the integral in powers of $(1/q^{2})$ and retain terms up to  $(1/q^{6})$ in this work. The first term in the expansion of expression (\ref{fig1a}) happens to be $1/q^{4}$. Neglecting the contributions of the twist-3 three-particle DA, $\Phi_{3\psi}^{p} (u)$=$h_{\psi}$  has been taken\cite{Ball:1999,Agaev:2011}. It is to be noted that, in vector meson dominance approximation, the factor $1/m_{\rho}^2$ is identified with the magnetic susceptibility of the quark condensate \cite{Ioffe:1984,Ball:2003}.
\par The Feynman diagram given by Fig. 1(b) contributes to the matrix element $T_{\mu\nu}$ through a bilocal tensor operator given by,
\begin{equation}\label{fig1b}
\fl T_{\mu\nu}^{(b)} (q,p)_{(tensor)} =\frac{8 g^2}{9}\sum_{\psi=u,d,s}e_{\psi}^2\frac{1}{m_{\psi}}\langle \overline{\psi}\psi\rangle \epsilon_{\mu\nu\alpha\beta}p^\alpha\frac{\partial}{\partial q_{\beta}} \int_{0}^{1} du \Phi_{3M}^{(\psi)\sigma}(u)\frac{1}{(q-up)^4}.
\end{equation}
In the considered approximation, $\Phi_{3\psi}^{\sigma} (u)$=$6u\overline{u} h_{\psi}$ \cite{Ball:1999,Agaev:2011}. The result of the integral is of the type $1/q^{6}$ in our approximation. The contributions from diagrams corresponding to Figs. 1(c) and 1(d) are obtained by expanding quark propagator close to the light-cone in a background gluon field \cite{Balitsky:1989}. Using equation of motion the covariant derivative of the gluon field strength tensor is converted to a quark-antiquark pair form. This leads to a bilocal tensor operator yielding

\begin{eqnarray}\label{eq44}
\fl T_{\mu\nu}^{(c+d)} (q,p)_{(tensor)}=\frac{g^2}{27}\sum_{\psi=u,d,s}e_{\psi}^2\frac{\langle \overline{\psi}\psi \rangle}
{m_{\psi}}\epsilon_{\mu\nu\alpha\beta}p^\alpha\frac{\partial}{\partial q_{\beta}} \int_{0}^{1}\int_{0}^{1} du dv\nonumber\\\times
(u\overline{u}-\frac{1}{2})\left[\frac{u}{q^4(1-uv)^2}+\frac{u}{(q^2 uv-p^2(1-uv))^2}\right]\Phi_{3M}^{(\psi)\sigma}(v).
\end{eqnarray}

\par Using light-cone expansion of the product of two currents \cite{Balitsky:1989} one gets diagrams (e) and (f) shown in Fig. 1. This results in a tensor-type of bilocal operator and the result can be expressed as
\begin{eqnarray}\label{eq45}
\fl T_{\mu\nu}^{(e+f)} (q,p)_{(tensor)}=\frac{g^2}{9}\sum_{\psi=u,d,s}e_{\psi}^{2}\frac{\langle\overline{\psi}\psi\rangle}{m_{\psi}} \epsilon_{\mu\rho\alpha\beta}p^\alpha  \int_{-1}^{1} du \int_{-1}^{u} dv\int_{0}^{1}dw
 \Phi_{3M}^{(\psi)\sigma}(w)\times \nonumber\\ \left[\frac{\overline{v}(g_{\nu}^{\beta}q^\rho+g_{\nu}^{\rho}q^\beta(1-(1+w+v\overline{w})/2))}{\{q^2\overline{v}\frac{\overline{w}}{2}-p^2(1+w+v\overline{w})/2\}^3}-\frac{(1+u)(g_{\nu}^{\beta}q^\rho+g_{\nu}^{\rho}q^\beta(1+w-\overline{w}u)/2)}{\{q^2(1+w-\overline{w}u)/2-p^2\overline{w}(1+u)/2\}^3}\right].
\end{eqnarray}
This is in contradiction to the result obtained in Ref \cite{Agaev:2011} where it has been found to vanish. The result of integral starts at $1/q^{4}$.
\par The diagrams represented by Figs. 1(e) and 1(f) also contribute to axial-type bilocal operator albeit with a linear quark mass term in the numerator. This can be retained for s-quark. The corresponding result is
\begin{eqnarray}\label{eq46}
\fl T_{\mu\nu}^{(e+f)} (q,p)_{(axial)}=\frac{-2}{3}g^2 e_{s}^2 m_{s} f_{M}^{(s)} \langle \overline{s}s\rangle\epsilon_{\mu\nu\alpha\beta}p^\beta\frac{\partial}{\partial q_{\alpha}} \int_{0}^{1} du u\overline{u}\int_{0}^{1} dv \Phi_{2M}^{(s)}(v)\nonumber\\ \times\left[\frac{1}{(q-uvp)^2}+\frac{1}{(q-(v+\overline{v}u)p)^2}\right].
\end{eqnarray}
This gives $1/q^{6}$-type contribution.
\par The expansion of quark propagator near light-cone in the background gluon field also gives rise to a contribution involving gluon DA, shown in Fig. 1(g), and is linear in quark mass:
\begin{eqnarray}\label{eq47}
\fl T_{\mu\nu}^{(g)} (q,p)=\frac{-g^2f_{M}^{0}}{24^2\sqrt{3}}e_{s}m_{s}\langle \overline{s}s\rangle \epsilon_{\mu\nu\alpha\beta}\overline{n}^{\beta}\overline{n}_{\lambda}p_\rho p_{\sigma}\frac{\partial}{\partial q_{\rho}}\frac{\partial}{\partial q_{\sigma}}\frac{\partial}{\partial q_{\lambda}}\frac{\partial}{\partial q_{\alpha}}\nonumber \\ \times  \int_{0}^{1} du\int_{0}^{u} dv\int_{0}^{1} dw\Phi_{2M}^{g}(w)(1-2\overline{u}-2v)(u-v)^2 \times\nonumber\\ \left[\frac{1}{(q+p(wu+\overline{w}v))^2}-\frac{1}{(q+p(w\overline{u}+\overline{w}\overline{v}))^2}\right],
\end{eqnarray}
where $\overline{n}$ is a light-cone constant vector \cite{Ali:2003,Kroll:2003}. We estimate this contribution to be of the order of $1/q^{8}$ and hence we drop it.
\par  We have also looked into the possibility of contribution from gluon condensate in place of quark condensate times quark mass. However, such contribution arising from light-cone expansion of quark propagator as well as from product of two currents near light cone, shown in Fig. 1(h), both vanish. Collecting all the contributions from Eqs. (\ref{fig1a}-\ref{eq47}) and retaining terms only upto order $(1/Q^6)$ in $F_{\gamma\gamma^{*}\rightarrow\eta^{(\prime)}}(Q^{2})$, we get
\begin{eqnarray}
\fl Q^2F_{\gamma\gamma^{*}\rightarrow \eta}(Q^2)=\frac{-16\pi\alpha_{s}}{81} \langle \overline{q}q \rangle \frac{1}{Q^4} \left[ \left\{\frac{5 h_{q}\cos\phi}{\sqrt{2}m_{q}} -\frac{\kappa_{s}}{m_{s}}h_{s}\sin\phi\right\} \right.\nonumber \\ \times \left \{\frac{Q^2}{m_{\eta}^{2}}- \frac{Q^2}{m_{\rho}^{2}}\log\left[\frac{Q^2}{m_{\eta}^{2}}\right]- \frac{19}{2}-\frac{2\pi^{2}}{3} + \frac{7}{2}\log\left[\frac{Q^2}{m_{\eta}^{2}}\right] \right.\nonumber\\ \left. + \left(-\frac{m_{\eta}^{2}}{m_{\rho}^2}+4\right) \log\left[\frac{Q^2}{m_{\eta}^{2}}\right]-\frac{3}{2}\log^{2}\left[\frac{Q^2}{m_{\eta}^{2}}\right] \right\} \nonumber\\ + \left(\frac{9m_{s}\kappa_{s}}{\sqrt{3}}\right) \left[\left(f_{\eta}^{0}-\sqrt{2}f_{\eta}^{8}\right)\left(46-\frac{4\pi^{2}}{3}
-14\log\left[\frac{Q^{2}}{m_{\eta}^{2}}\right]+2\log^{2}\left[\frac{Q^{2}}{m_{\eta}^{2}}\right]\right) \right. \nonumber\\ + \left\{f_{\eta}^{0}\left(6B_{2}^{q}(\eta_{0})L^{\frac{48}{75}}-\frac{B_{2}^{g}}{17}L^{\frac{107}{75}}\right) -6\sqrt{2}f_{\eta}^{8}B_{2}^{q}(\eta_{8})L^{\frac{2}{3}}\right\}\nonumber\\ \left. \left. \times \left\{\frac{-316}{9}+2\pi^{2}+\frac{31}{6}
\log\left[\frac{Q^{2}}{m_{\eta}^{2}}\right]-\frac{1}{2}\log^{2}\left[\frac{Q^{2}}{m_{\eta}^{2}}\right]\right\}\right]\right], \label{eq48}
\end{eqnarray}
where $\kappa_{s}$=
$\langle \overline{s}s\rangle/\langle\overline{q}q\rangle$. For $Q^2F_{\gamma\gamma^*\rightarrow\eta^{\prime}}(Q^2)$, one has to make the substitution $(\cos\phi\rightarrow\sin\phi,-\sin\phi\rightarrow\cos\phi,m_{\eta}\rightarrow m_{\eta^{\prime}},f_{\eta}^{0,8}\rightarrow f_{\eta^{\prime}}^{0,8})$ in the above equation. Thus, twist-6 contributions produce $1/Q^{4}$ correction \cite{Agaev:2011}, like twist-4 correction but with smaller coefficients.
\section{Numerical analysis of result}
In this work, we use two loop result for running QCD coupling constant with $\Lambda_{QCD}^{(4)}$=326 MeV and four active flavors. In addition, we use following constants at renormalization scale $\mu_{0}$=1 GeV\cite{Kroll:2003,Charng:2006,Ball:2007,Singh:2013,Singh:2009,Agaev:2010,Agaev:2014}:
\par $\langle \overline{q}q\rangle$= -$(0.240\pm0.010$ GeV$)^{3}$, $\kappa_{s}$=$(0.8\pm0.1)$,
$m_{q}$=$(4.5\pm0.5)$MeV, $m_{s}$=$(100\pm10)$MeV,$\phi$=$39.3\degree\pm1.0\degree$, $f_{q}$=$(1.07\pm0.02)f_{\pi}$, $f_{s}$=$(1.34\pm0.06)f_{\pi}$, $B_{2}^{q}(\eta_{0})$=$B_{2}^{q}(\eta_{8})$=$0.115\pm0.035$, $B_{2,g}$=$18\pm2$, $h_{q}$=$0.0015\pm0.004$ $GeV^{3}$, $h_{s}$=$0.087\pm0.006$ $GeV^{3}$ and $m_{\rho}$=$0.77$ GeV with zero width $\Gamma_{\rho}$=0.
Since in this work we are calculating only twist-6 corrections to TFFs of $\eta$ and $\eta^{\prime}$ mesons, we shall be using results on leading order and next-to-leading order  power corrections arising from lower twists from existing literature. In Table. 1, we have displayed the coefficients of $1/Q^{4}$  and $1/Q^{6}$ in $F_{\gamma\gamma^{*}\rightarrow\eta^{(\prime)}}(Q^{2})$ for $Q^{2}$ = 5, 10, 50 GeV$^{2}$. In Table. 2, we have displayed the composition of our result for TFFs as ratios of different contributing Lorentz structures to the total twist-6 result. It demonstrates as to how the cancellation among the pseudoscalar and tensor structures results in a small value of overall result. Since $h_{q}$ has been found to influence the result on twist-6 contribution considerably, it is worth discussing the origin of its uncertainties. If the errors of $f_{q,s}$ and $\phi$ are treated as uncorrelated, then expression(\ref{heta1}) gives $h_{q}$=$(0.0015\pm0.004)$ $GeV^{3}$ \cite{Beneke:2003}, $\sim 270\%$ uncertainty. However, QCD sum rule estimate by one of us yields $h_{q}$=$(0.0025\pm0.0009)$ $GeV^{3}$ \cite{Singh:2013}. In Ref. \cite{Ball:2007}, the authors have considered $\frac{h_{q}}{(2m_{q})}$ which normalizes twist-3 DAs of $\eta_{q}$. Working to leading order in chiral expansion they set $\frac{h_{q}}{(2m_{q})}$=$f_{q}B_{0}$ with $B_{0}$= $m_{\pi}^{2}/(2m_{q})$=$-2\langle 0\mid\overline{q}q\mid 0\rangle/f_{\pi}^{2}$. With the uncertainty in $\langle\overline{q}q\rangle$ and $f_{q}$ as given above  and $m_{q}$=(4-5) MeV, one finds $h_{q}$=$(0.0021\pm0.0005)$ $GeV^{3}$. Another approximate numerical estimate of $h_{q}$ can be given using the octet-singlet basis of $\eta-\eta^{\prime}$ system. In Ref. \cite{Kim:2008}, $\mu_{\eta}$=$3m_{\eta}^{2}/(m_{u}+m_{d}+4m_{s})$, chirally enhanced factor, has been introduced in the matrix element of octet pseudoscalar quark current between vacuum and one $\eta$-state, assuming it to be a pure octet, in line with $\mu_{\pi}$ and $\mu_{K}$. Similarly, in Ref. \cite{Ali:2003}, $\mu_{\eta^{\prime}}$=$3m_{\eta^{\prime}}^{2}/(2m_{u}+2m_{d}+2m_{s})$ has been introduced in the matrix element of singlet pseudoscalar quark current between vacuum and one $\eta^{'}$- state assuming it to be a pure singlet state. Writing these matrix elements in both quark-flavor as well as octet-singlet basis one gets
\begin{equation}
\frac{h_{q}\cos\phi}{\sqrt{12}m_{q}}+\frac{h_{s}\sin\phi}{\sqrt{6}m_{s}}=f_{\eta}^{8}\mu_{\eta},
\end{equation}
\begin{equation}
\frac{h_{q}\sin\phi}{\sqrt{6}m_{q}}+\frac{h_{s}\cos\phi}{\sqrt{12}m_{s}}=f_{\eta^{\prime}}^{0}\mu_{\eta^{\prime}}.
\end{equation}
These equations can be solved together to estimate $h_{q}$ in an approximate way since $\eta$ and $\eta^{\prime}$ are not simply octet and singlet states respectively as treated in this derivation. Taking the numerical values of $f_{\eta}^{8}$ and $f_{\eta^{\prime}}^{0}$, as well as (for the sake of consistency) the quark masses from Ref.\cite{Singh:2013}, we estimate $h_{q}$ $\simeq 0.0057 GeV^{3}$. This barely touches the upper limit of the first estimate given in this paragraph.

In Figs. 2 and 7, we have shown our result for twist-6 correction to the TFF $F_{\gamma\gamma^*\rightarrow\eta^{(\prime)}}(Q^2)$ for $h_{q}$=0.0055, 0.0015 and -0.0025 which covers the range obtained in the first estimate. In Tables. 1 and 2 as well as later in Figs. 3 and 6, we have picked up one $h_{q}$ $\simeq 0.0020 GeV^{3}$ which lies in more narrow ranges given in the second and third estimates. For the argument of the running QCD coupling constant, we use frequently used scale $\mu^{2}$=$Q^{2}$. In Figs. 3 and 6, we have compared our result for twist-6 correction to $F_{\gamma\gamma^{*}\rightarrow\eta^{(\prime)}}(Q^{2})$ with the result for twist-4 correction to the same from Ref. \cite{Agaev:2014}. In Figs. 4 and 8, we superimpose our results of twist-6 corrections on those obtained in Ref. \cite{Kroll:2013} for TFFs to leading twist accuracy and NLO of perturbative QCD for a specific set of parameters (Gegenbauer coefficients of order 2) obtained from fitting the data. The shaded area shows the uncertainty in our result due to uncertainty in  various input parameters, as given above in this section, and has been shown separately in  Figs. 4(a) and 8(a) for clarity.

\begin{table}
\begin{center}
\caption{Coefficients of $\frac{1}{Q^{4}}$ and  $\frac{1}{Q^{6}}$ in our result for the TFF for $F_{\gamma\gamma^{*}\rightarrow\eta^{(\prime)}}(Q^{2})$ at different momenta. Parameters used for evaluation: $\langle\overline{q}q\rangle=(-0.24)^{3}  GeV^{3}, \kappa_{s}=\frac{\langle\overline{s}s\rangle}{\langle\overline{q}q\rangle}=0.8, \phi=40.3\degree , h_{q}=0.0020 GeV^{3}, h_{s}= 0.087 GeV^{3}, f_{q}=1.07 f_{\pi}, f_{s}=1.34 f_{\pi}, m_{q}=4.5  MeV, m_{s}=100  MeV, B_{2}^{q}(\eta_{0})=B_{2}^{q}(\eta_{8})=0.15, B_{2}^{g}=16$.} \label{tab:twist6}
\begin{tabular}{ccccl}
\hline\hline
\multirow{2}{*}{$Q^2$} & \multicolumn{2}{c}{Coefficient of $\frac{1}{Q^4}$} & \multicolumn{2}{c}{Coefficient of $\frac{1}{Q^6}$} \\
& $\eta$ & $\eta^{\prime}$ & $\eta$ & $\eta^{\prime}$   \\
\hline
5 	& -0.0048 & -0.0103  & -0.0220  & -0.0287 \\
10	& -0.0075 & -0.0148 & -0.0266 & -0.0357 \\
50	& -0.0118 & -0.0218 & -0.0445 & -0.0580\\
\hline\hline
\end{tabular}
\end{center}
\end{table}

\begin{figure}[H]
\includegraphics[width=0.5\textwidth]{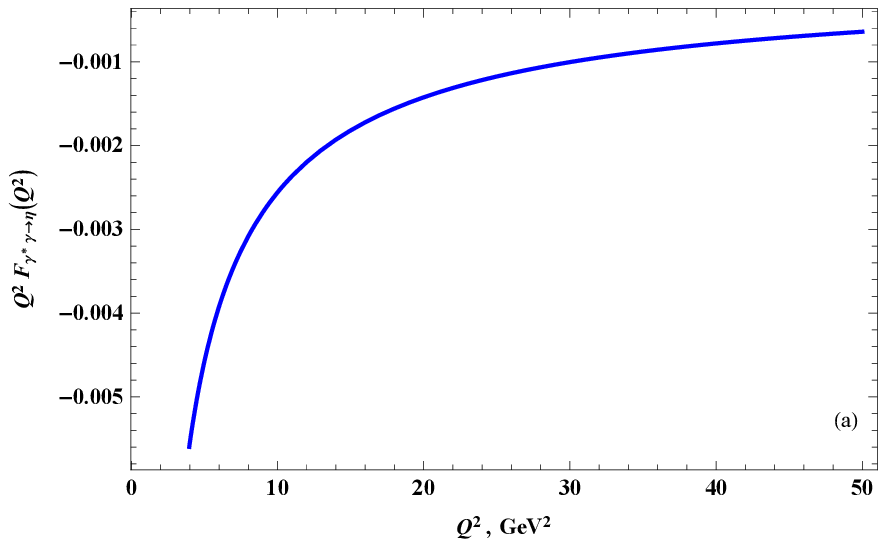}
\hfill\includegraphics[width=0.5\textwidth]{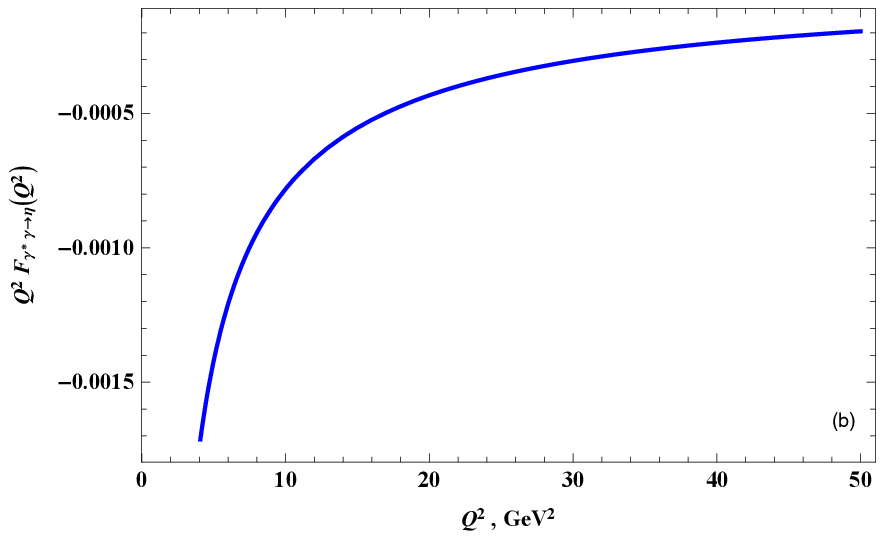}\\\\
\includegraphics[width=0.5\textwidth]{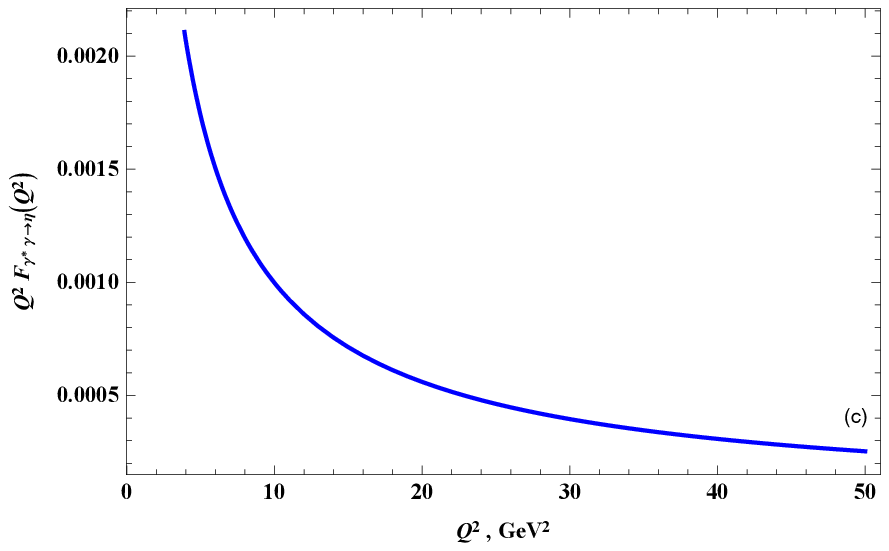}
\caption{Plot of our result for twist-six correction to the TFF $F_{\gamma\gamma^{*}\rightarrow\eta}(Q^{2})$ for $h_{q}$=0.0055 (a), 0.0015 (b) and -0.0025 (c).}
\end{figure}
\begin{figure}[H]
\begin{center}
\includegraphics[width=0.55\textwidth]{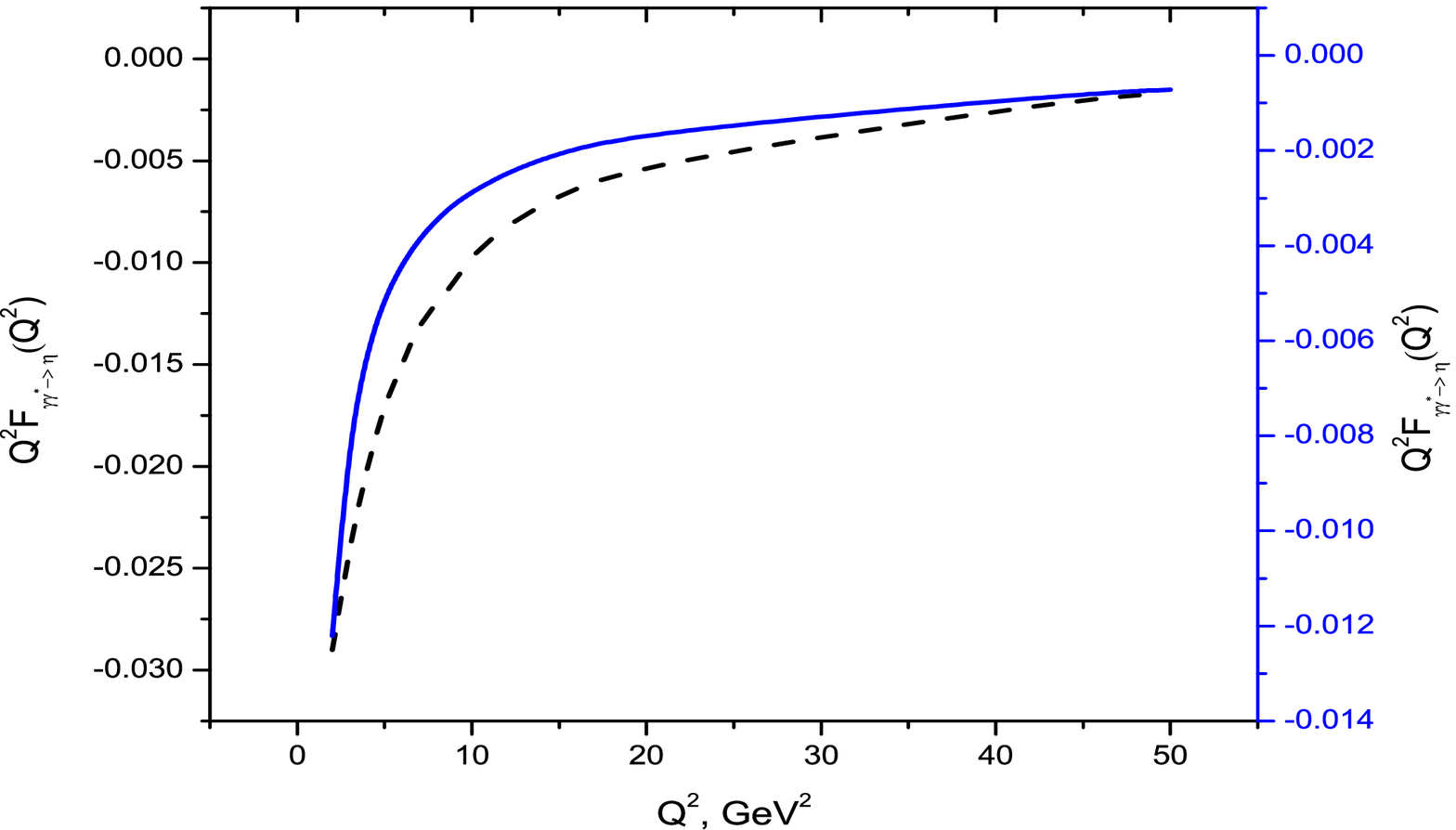}
\caption{Comparison of our result for twist-six correction to the TFF $F_{\gamma\gamma^{*}\rightarrow\eta}(Q^{2})$ (solid line, right scale) with twist-four correction \cite{Agaev:2014} to the same (dashed line, left scale). Parameters used are the same as used in Table 1.}
\end{center}
\end{figure}
\begin{table*}[h]
\begin{center}
\caption{Ratios of different Lorentz structures to the total twist-six result for the TFF for $F_{\gamma\gamma^*\rightarrow\eta^{(\prime)}}(Q^2)$ at different momenta. Parameters used for this evaluation are the same as those used in Table.1. } \label{tab:twist61}
\begin{tabular}{rllllll}
\hline\hline
\multirow{2}{*}{$Q^2$} & \multicolumn{3}{c}{$\eta$} & \multicolumn{3}{c}{$\eta^{\prime}$} \\
& $\frac{F_{pseudo-6}}{F_{twist-6}}$ & $\frac{F_{axial-6}}{F_{twist-6}}$  & $\frac{F_{tensor-6}}{F_{twist-6}}$    & $\frac{F_{pseudo-6}}{F_{twist-6}}$ & $\frac{F_{axial-6}}{F_{twist-6}}$  &  $\frac{F_{tensor-6}}{F_{twist-6}}$   \\
\hline
 5	& 1.850	& 0.020    & -0.870  & 1.230   & -0.041   & -0.189 \\
10	& 1.742 & 0.010   & -0.752   & 1.205   & -0.011   & -0.194 \\
50  & 1.524 & 0.002   & -0.526    & 1.153   & -0.001   & -0.152 \\
\hline\hline
\end{tabular}
\end{center}
\end{table*}
\begin{figure}[H]
\includegraphics[width=0.5\textwidth]{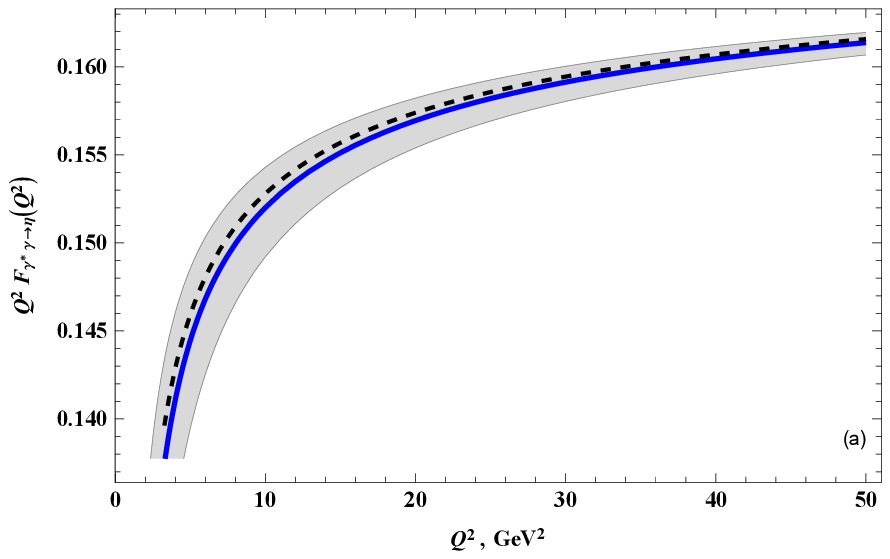}
\hfill\includegraphics[width=0.5\textwidth]{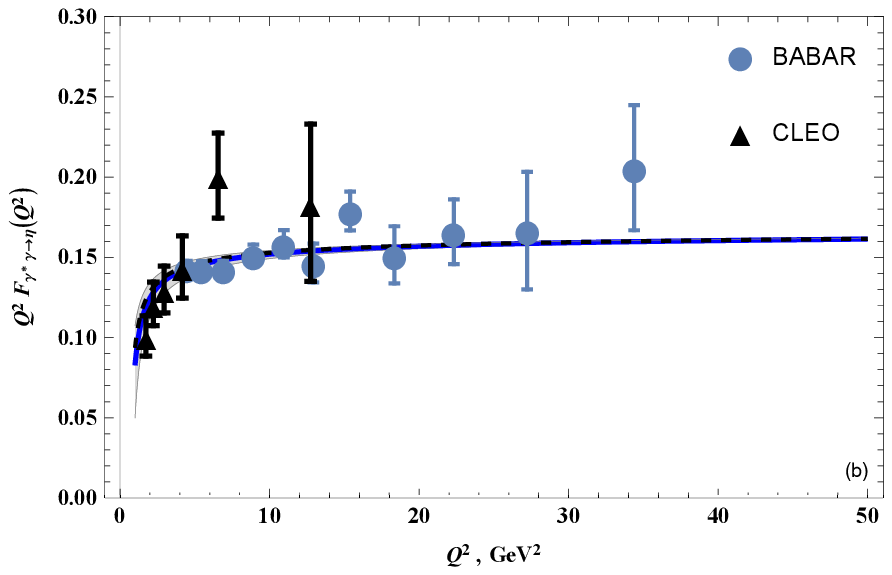}
\caption{Plot of our result for twist-six corrections (solid lines) superimposed on the result obtained in Ref.\cite{Kroll:2013} (a)(dashed line, plotted by us) for the TFF $F_{\gamma\gamma^{*}\rightarrow\eta}(Q^{2})$. The shaded area corresponds to the uncertainty in our result due to variation in the input parameters as given in the text. In (b) the same results are compared with data from Refs. \cite{Gronberg:1998,Sanchez:2011}.}
\end{figure}
\begin{figure}[H]
\includegraphics[width=0.5\textwidth]{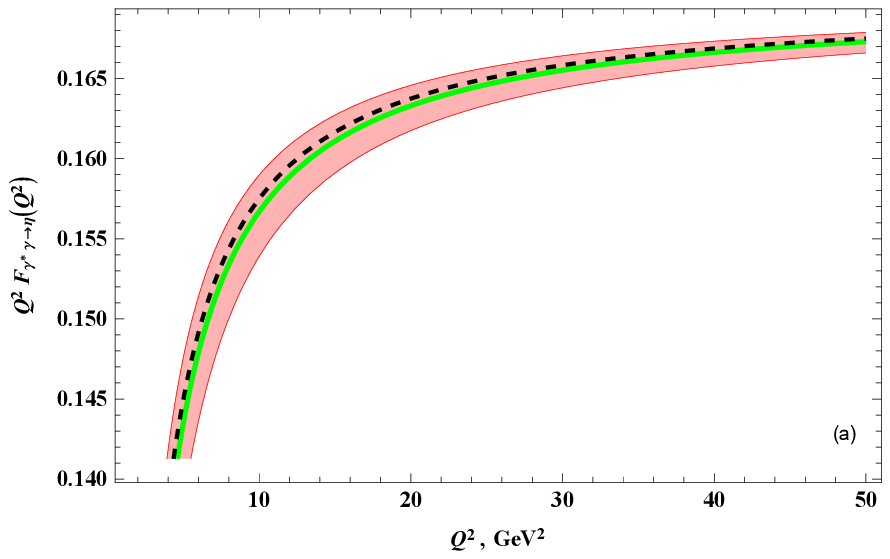}
\hfill\includegraphics[width=0.5\textwidth]{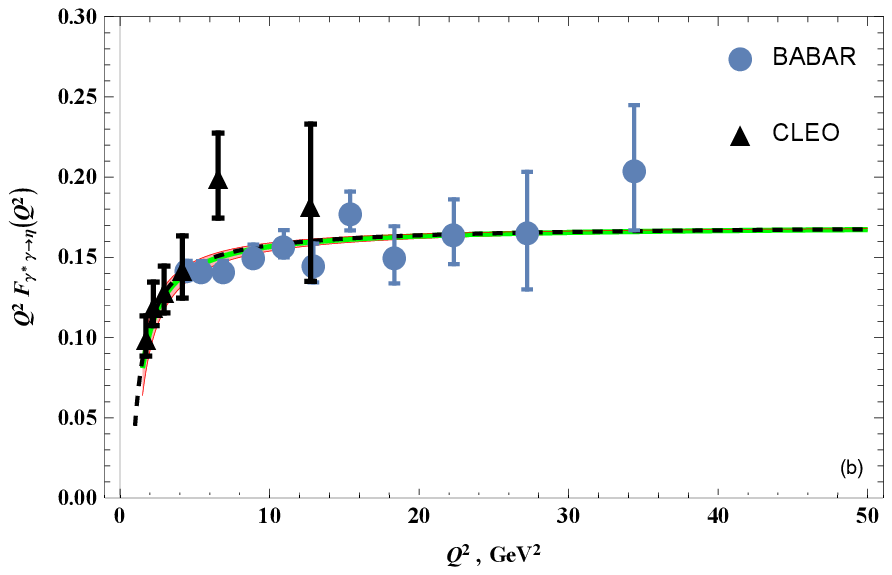}\\
\caption{Plot of our result for twist-six corrections (solid lines) superimposed on the result obtained in Ref.\cite{Agaev:2014} (a)(dashed line, plotted by us) for the TFF $F_{\gamma\gamma^{*}\rightarrow\eta}(Q^{2})$. The shaded area corresponds to the uncertainty in our result due to variation in the input parameters as given in the text. In (b) the same results are compared with the data from Refs. \cite{Gronberg:1998,Sanchez:2011}.}
\end{figure}
In Figs. 4(b) and 8(b), we show this combination of results alongwith the data from Refs.\cite{Gronberg:1998,Sanchez:2011}. Results on TFFs for $\eta$ and $\eta^{\prime}$ mesons obtained in Ref \cite{Agaev:2014} include NLO analysis of perturbative corrections, charm-quark contribution, SU(3)-flavor breaking effects and the axial anomaly contributions to the power-suppressed twist-4 DAs. In Figs. 5(a) and 9(a) we have shown the results obtained by superimposing our results to those obtained in Ref. \cite{Agaev:2014} for a specific set of parameters. Again the shaded area shows the uncertainty in our results only due to uncertainty in various input parameters as given above in this section. Figs. 5(b) and 9(b) show the comparison of this combination with data points.
\begin{figure}[H]
\includegraphics[width=0.55\textwidth]{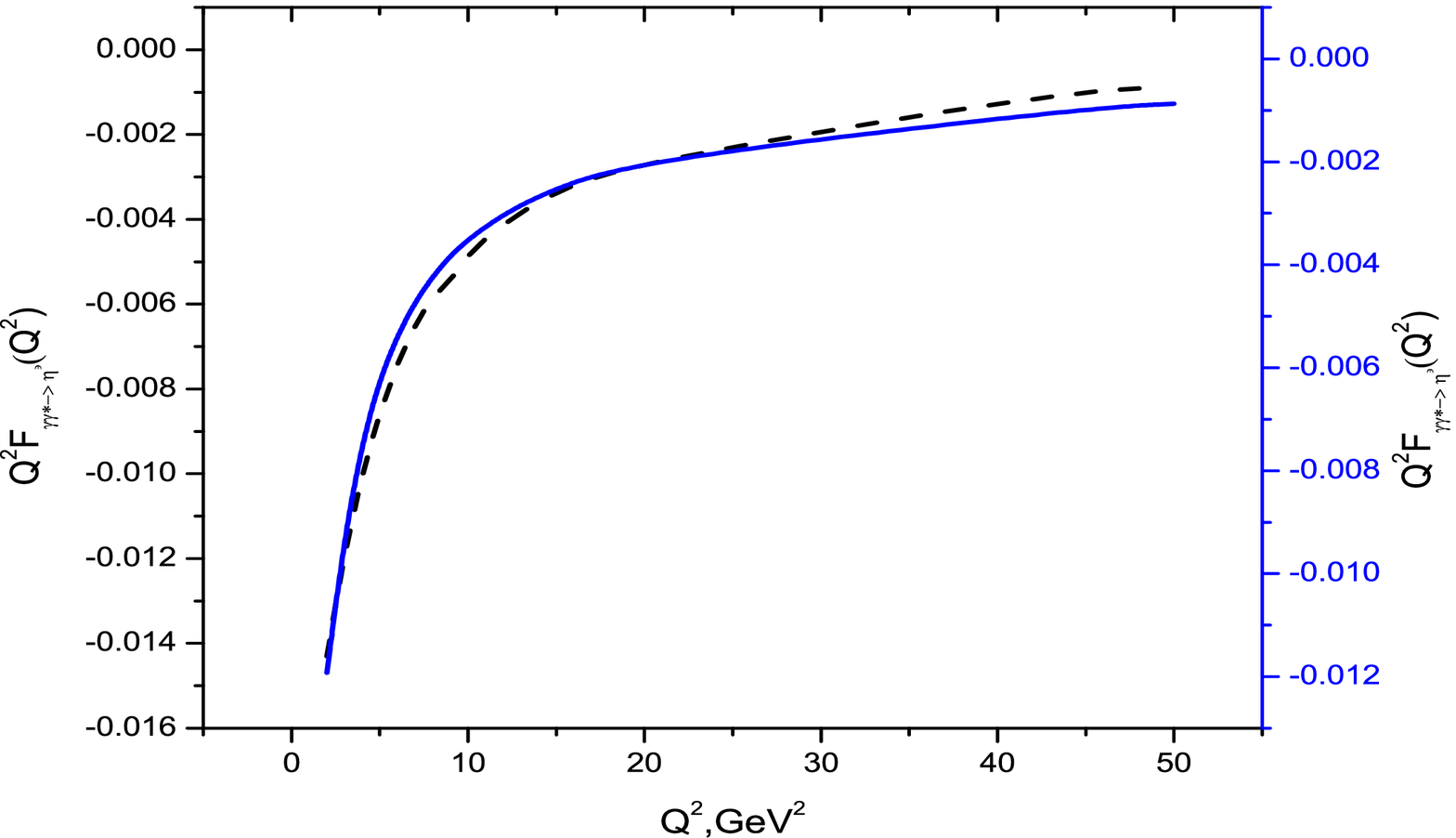}
\caption{Comparison of our result for twist-six correction to the TFF $F_{\gamma\gamma^{*}\rightarrow\eta^{\prime}}(Q^{2})$ (solid line, right scale) with twist-four correction \cite{Agaev:2014} to the same (dashed line, left scale). Parameters used are the same as used in Table 1.}
\end{figure}
\begin{figure}[H]
\includegraphics[width=0.5\textwidth]{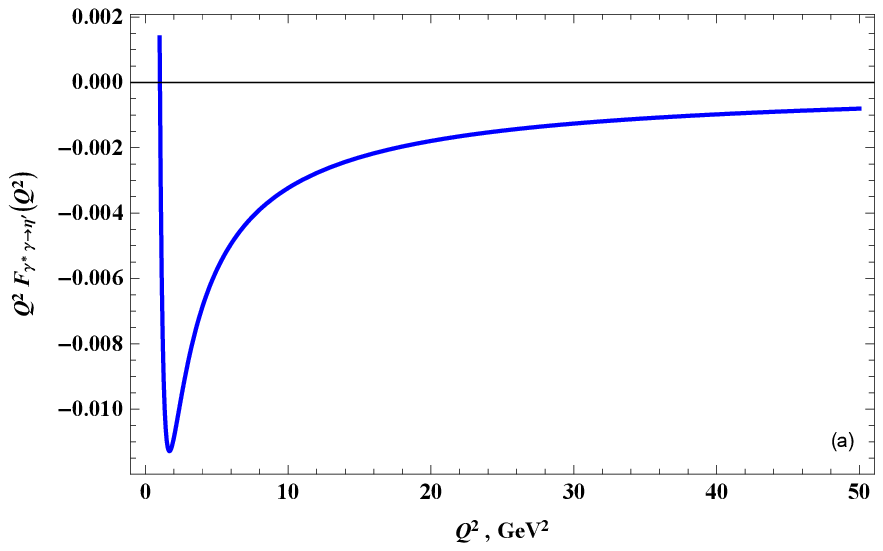}
\hfill\includegraphics[width=0.5\textwidth]{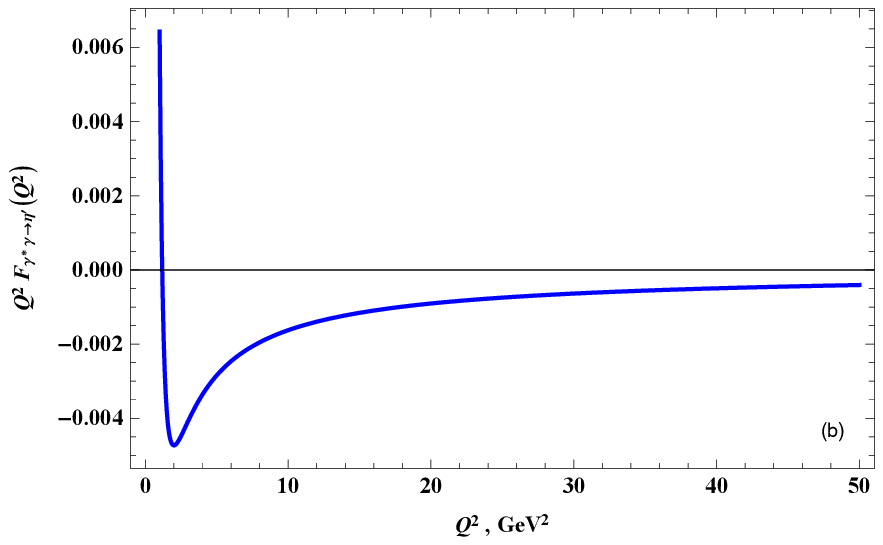}\\\\
\includegraphics[width=0.5\textwidth]{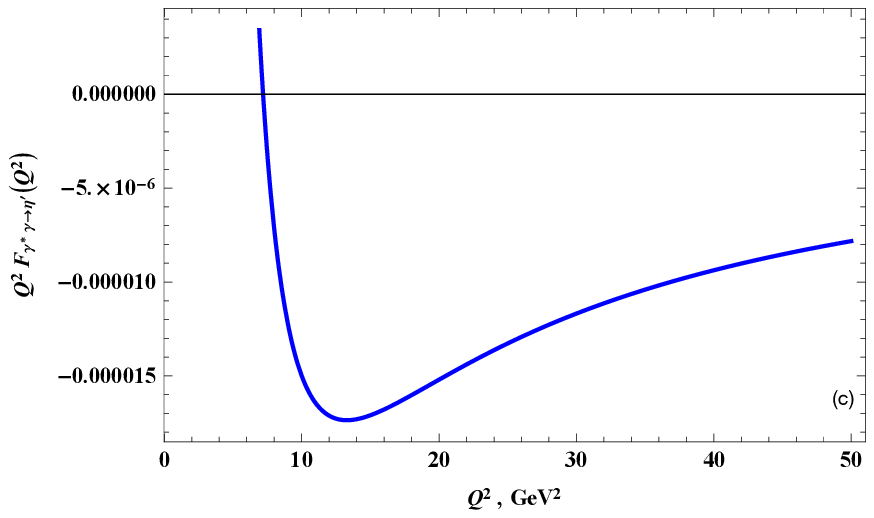}
\caption{Plot of our result for twist-six correction to the TFF $F_{\gamma\gamma^{*}\rightarrow\eta^{\prime}}(Q^{2})$ for $h_{q}$=0.0055 (a), 0.0015 (b) and -0.0025 (c).}
\end{figure}
\begin{figure}[H]
\includegraphics[width=0.5\textwidth]{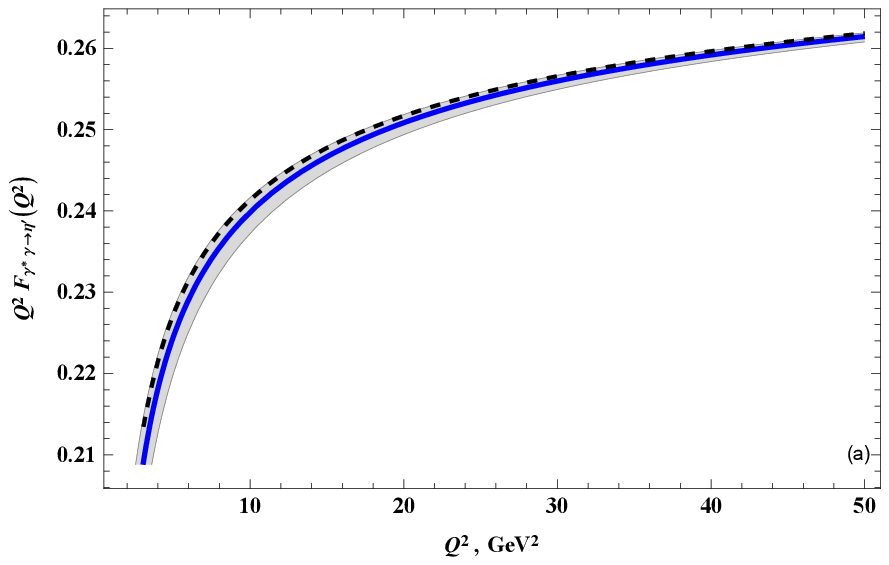}
\hfill\includegraphics[width=0.5\textwidth]{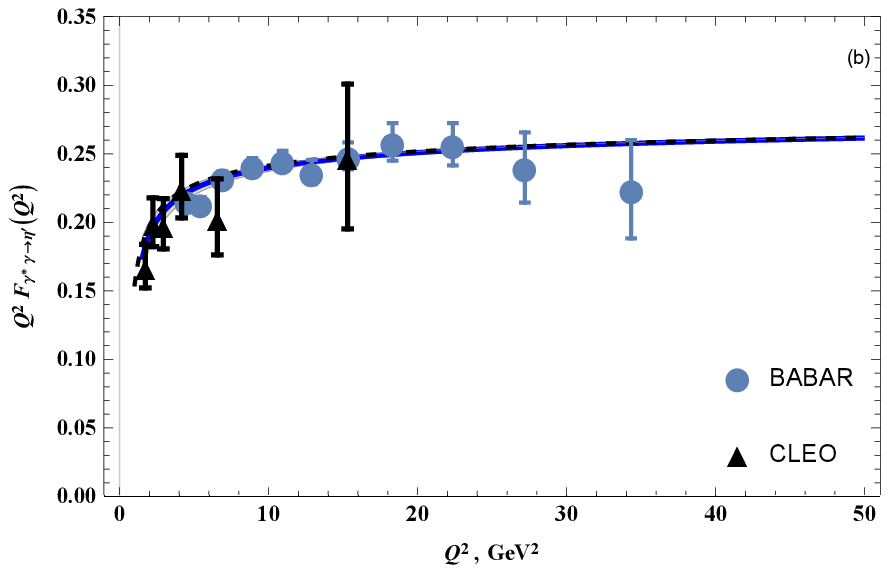}\\
\caption{Plot of our result for twist-six corrections (solid lines) superimposed on the result obtained in Ref.\cite{Kroll:2013}(a) (dashed line, plotted by us) for the TFF $F_{\gamma\gamma^{*}\rightarrow\eta^{\prime}}(Q^{2})$. The shaded area corresponds to the uncertainty in our result due to variation in the input parameters as given in the text. In (b) the same results are compared with the data from Refs. \cite{Gronberg:1998,Sanchez:2011}.}
\end{figure}
\begin{figure}[H]
\includegraphics[width=0.5\textwidth]{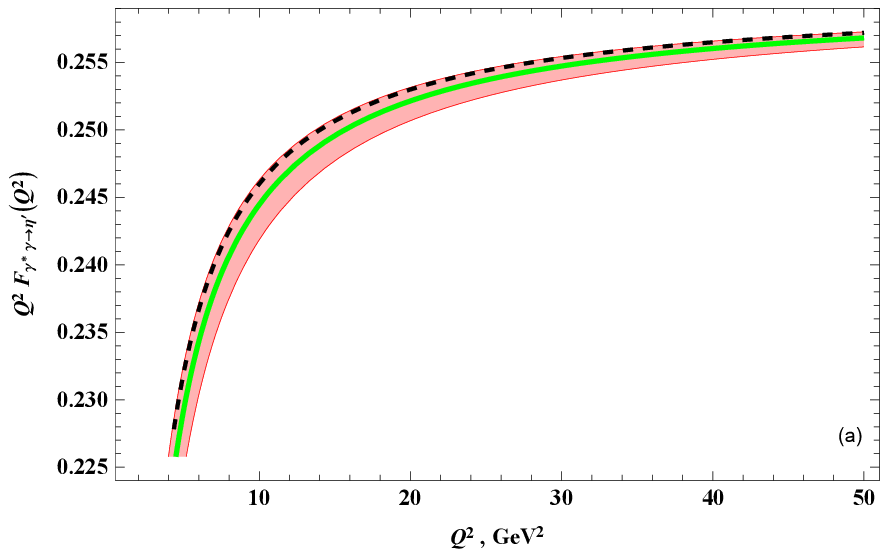}
\hfill\includegraphics[width=0.5\textwidth]{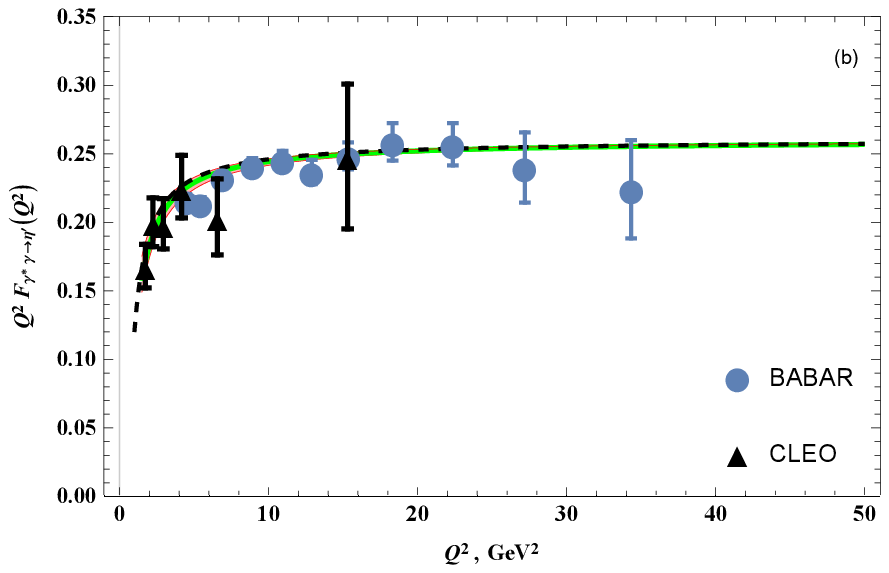}\\
\caption{Plot of our result for twist-six corrections (solid lines) superimposed on the result obtained in Ref.\cite{Agaev:2014}(a) (dashed line, plotted by us) for the TFF $F_{\gamma\gamma^{*}\rightarrow\eta^{\prime}}(Q^{2})$. The shaded area corresponds to the uncertainty in our result due to variation in the input parameters as given in the text. In (b) the same results are compared with the data from Refs. \cite{Gronberg:1998,Sanchez:2011}.}
\end{figure}

\section{Summary and conclusion}
Our result on twist-6 corrections to TFFs for $\eta$ and $\eta^{\prime}$ mesons starts with $1/Q^{4}$, as is the case for twist-4 corrections \cite{Agaev:2014}, but with a smaller coefficient. For this calculation we have considered a subset of twist-6 operators which can be factorized as a product of two gauge invariant twist-3 operators or one twist-2 and one twist-4 operator. Our general framework is to use light-cone expansion of product of two currents or of a quark propagator but we have also used additional Feynman diagrams which are not accounted for by light-cone expansion. We found that while gluon condensate along with twist-2 DA does not contribute to TFFs, the meson DA with two valance gluons, being a special case for $\eta$ and $\eta^{\prime}$ mesons, contributes, has a higher order term in momentum expansion and hence, has been dropped in our approximation. Nevertheless, gluon DA contribution appears due to quark-gluon mixing and renormalization group evolution. Non-factorizable operators are expected to give negligible contribution \cite{Braun:2000}. We have used vector-dominance model to regulate the result when one of the quark lines goes to the mass-shell. Twist expansion provides a systematic way to calculate higher order power corrections to exclusive processes, our endeavour is to estimate contribution arising from twist-6 operators to TFFs of $\eta$ and $\eta^{\prime}$ mesons. Since the mesons involved are not simply Goldstone bosons due to their anomalous masses, we have included their masses as well as the s-quark mass (linear) in our result.  We found that $h_{q}$, which is the first term in Gegenbauer expansion of twist-3 DA of pseudoscalar-type, introduces considerable uncertainty if it is not well constrained. We observed that twist-6 contribution is a couple of times smaller in magnitude than its counterpart for twist-four for the $\eta$ meson, but for $\eta^{\prime}$ meson these two contributions are comparable. As far as light-cone sum rules are concerned, we found that they introduce less than $10\%$ modification for twist-4 operators for $Q^{2}>7 GeV^{2}$ in \cite{Agaev:2014} and around $25\%$ modification for twist-6 operators in \cite{Agaev:2011}. In our case they may modify the result by up to $20\%$  due to higher Borel mass and continuum threshold. This will be a small change to the total result for TFFs $F_{\gamma\gamma^{*}\rightarrow\eta^{(\prime)}}(Q^{2})$. We feel that any further higher twist correction will make insignificant improvement in the total result for TFFs. Constraining parameters such as $h_{q}$ is better called for. Including few more terms in the expansion of lower-twist DAs, taking non-valance quark(gluon) contribution and taking into account the $k_{T}$-corrections when $Q^2\sim$ a few $GeV^2$ are other steps which can be taken to improve theoretical results in this approach.

\ack Authors gratefully acknowledge partial financial support from DST-SERB. SDP also acknowledges partial financial support from UGC-BSR. JPS thanks S. J. Brodsky for critical comments and suggestions.
\appendix
\section{Decay constants, mixing in twist-2 DAs , anomaly equation and three-particle twist-3 DAs}
Here we give some more details of the quantities that have been used in the main text.
\subsection{Meson decay constants and mixing in twist-2 DAs}
Decay constants $f_{M}^{(i)}$ are parameterized in terms of two basic decay constants and two mixing angles  \cite{Feldmann:1999,Feldmann:2000}.
\begin{equation}
\fl \left(\begin{array}{cc} f_\eta^{(q)} & f_\eta^{(s)}\\
f_{\eta'}^{(q)} & f_{\eta'}^{(s)} \end{array}\right)
=\left(\begin{array}{cc} cos\theta_{q} & -sin\theta_{s}\\
sin\theta_{q} & cos\theta_{s} \end{array}\right)
\left(\begin{array}{cc} f_q & 0\\
0 & f_s \end{array}\right).
\end{equation}
It is well known that axial current has anomalaus divergence:
\begin{equation}
\fl \partial^{\mu}(\overline{\psi}\gamma_{\mu}\gamma_{5}\psi)=2i  m_{\psi}\overline{\psi}\gamma_{5}\psi-\frac{\alpha_{s}}{4\pi}G_{\mu\nu}^a \tilde{G}^{a\mu\nu}, (\psi=u,d,s),
\end{equation}
\begin{equation}
\fl \tilde{G}^{a\mu\nu}=\frac{1}{2}\epsilon^{\mu\nu\rho\sigma}G_{\rho\sigma}^{a}, \epsilon^{0123}=+1.
\end{equation}
 This $U(1)_{A}$ anomaly leads to significant mixing between (u,d) and s-quark states. The difference $|\theta_q -\theta_s|$ is determined by OZI- rule violating contribution and not by $SU(3)_{F}$ breaking. Phenomenological analysis gives $|\frac{\theta_q -\theta_s}{\theta_q +\theta_s}|\ll 1$. Hence, it is common to use $\theta_{q}\simeq\theta_{s}\simeq\phi$ and parameterize the mixing with the matrix $U(\phi)$ given by \cite{Feldmann:2000},
\begin{equation}
\fl U(\phi)=\left(
           \begin{array}{cc}
           \cos\phi & -\sin\phi \\
             \sin\phi & \cos\phi\\
           \end{array}
         \right).
\end{equation}
In QF basis one introduces pure non-strange and strange states $\mid\eta_{q}\rangle=\frac{1}{\sqrt{2}}\mid(u\overline{u}+d\overline{d})\rangle$ and $|\eta_{s}\rangle =\mid s\overline{s}\rangle$ leading to,
\begin{equation}
\fl \left(
  \begin{array}{c}
    \eta \\
    \eta^{\prime} \\
  \end{array}
\right)=U(\phi)\left(
                 \begin{array}{c}
                   \eta_{q} \\
                   \eta_{s} \\
                 \end{array}
               \right).
\end{equation}

Because the gluon and flavor singlet quark DAs mix under evolution, the gluon DA is usually assigned the same factor as for flavor singlet quark DA. The mixing equation has a $2\times2$ matrix form which has been solved \cite{Ali:2003,Agaev:2004,Agaev:2010}. Keeping only the first non-asymptotic term in the singlet component of the twist-2 DAs, the quark-antiquark and gluon DAs can be written as
\begin{equation}
\fl \Phi_{\eta_{0}}^{(q)}(u,\mu^2)=6u\overline{u}[1+A(\mu^2)-5A(\mu^2)u\overline{u}],
\end{equation}
\begin{equation}
\fl \Phi_{\eta_{0}}^{(g)}(u,\mu^2)=u^{2}\overline{u}^{2}(u-\overline{u})B(\mu^2).
\end{equation}
For $n_{f}$=4, the functions $A(\mu^2)$ and $B(\mu^2)$ are given by \cite{Agaev:2004,Agaev:2010},
\begin{equation}
\fl A(\mu^2)=6 B_{2}^{q}L^{\frac{48}{75}}(\mu^2)-\frac{B_{2}^{g}}{17}L^{\frac{107}{75}}(\mu^2),
\end{equation}
\begin{equation}
\fl B(\mu^2)=19 B_{2}^{q}L^{\frac{48}{75}}(\mu^2)+5B_{2}^{g}L^{\frac{107}{75}}(\mu^2),
\end{equation}
where $\mu_{0}^{2}=1  GeV^2 $ is a normalization scale and
\begin{equation}
\fl L(\mu^2)=\frac{\alpha_{s}(\mu^2)}{\alpha_{s}(\mu_{0}^{2})}.
\end{equation}
The DA of octet state has only the quark component $\Phi_{\eta 8}(u,\mu^2)$ where $C(\mu^2)$ replaces $A(\mu^2)$:
\begin{equation}
\fl C(\mu^2)=6B_{2}^{q}L^{2/3}.
\end{equation}

\par In the singlet-octet basis, the quark currents are written as
\begin{eqnarray}
\fl J_{\mu5}^{8}&=&\frac{1}{\sqrt{6}}(\overline{u}\gamma_{\mu}\gamma_{5}u+\overline{d}\gamma_{\mu}\gamma_{5}d-2\overline{s}\gamma_{\mu}\gamma_{5}s),
\\
\fl J_{\mu5}^{0}&=&\frac{1}{\sqrt{3}}(\overline{u}\gamma_{\mu}\gamma_{5}u+\overline{d}\gamma_{\mu}\gamma_{5}d+\overline{s}\gamma_{\mu}\gamma_{5}s).
\end{eqnarray}

The corresponding decay constants are defined as,
\begin{equation}
\fl \langle M(p)\mid J_{\mu5}^{a}\mid0\rangle = -i f_{M}^{a}p_{\mu}; a=8,0.
\end{equation}
These decay constants are related to those in QF basis as follows,
\begin{eqnarray}
\fl f_{\eta}^{(q)}&=&f_{q}\cos\phi=\frac{1}{\sqrt{3}}(\sqrt{2}f_{\eta}^{0}+f_{\eta}^{8}),
\\
\fl f_{\eta}^{(s)}&=&-f_{s}\sin\phi=\frac{1}{\sqrt{3}}(f_{\eta}^{0}-\sqrt{2}f_{\eta}^{8}),
\end{eqnarray}
\begin{eqnarray}
\fl f_{\eta^{\prime}}^{(q)}&=&f_{q}\sin\phi=\frac{1}{\sqrt{3}}(\sqrt{2}f_{\eta^{\prime}}^{0}+f_{\eta^{\prime}}^{8}),
\\
\fl f_{\eta^{\prime}}^{(s)}&=&f_{s}\cos\phi=\frac{1}{\sqrt{3}}(f_{\eta^{\prime}}^{0}-\sqrt{2}f_{\eta^{\prime}}^{8}).
\end{eqnarray}
\subsection{Matrix elements of anomaly equation and three-particle twist-3 DAs}
Introducing,
\begin{eqnarray}
\fl a_{M}&=&\langle M(p)\mid\frac{\alpha_{s}}{4\pi}G_{\mu\nu}^a \tilde{G}^{a\mu\nu}\mid0\rangle,
\\
\fl h_{M}^{q}&=&2i m_{q}\langle M(p)\mid\frac{1}{\sqrt{2}}(\overline{u}\gamma_{5}u+\overline{d}\gamma_{5}d)|0\rangle,
\\
\fl h_{M}^{s}&=&2i  m_{s}\langle M(p)\mid\overline{s}\gamma_{5}s|0\rangle,
\end{eqnarray}
one finds,
\begin{equation}\label{eq15}
\fl a_{M}=\frac{1}{\sqrt{2}}(h_{M}^{q}-f_{M}^{q}m_{M}^{2})=h_{M}^{s}-f_{M}^{s}m_{M}^{2},
\end{equation}
where  $ m_{q}= (m_{u}+m_{d})/2$.
\\The three particle twist-3 DA is defined as \cite{Belyaev:1995, Duplancic:2015},
 \begin{eqnarray}
 \fl \langle M(p)\mid \overline{r}(x)g G_{\mu\nu}^{n}(vx)\frac{\lambda^{n}}{2}\sigma _{\alpha\beta}\gamma_{5} r(0)\mid 0\rangle = i f_{3r}[(p_{\mu}p_{\alpha}g_{\nu\beta}-p_{\nu}p_{\alpha}g_{\mu\beta})-(p_{\mu}p_{\beta}g_{\nu\alpha}-p_{\nu}p_{\beta}g_{\mu\alpha})]\times\nonumber\\\int_{0}^{1}d\alpha_{1}d\alpha_{2}d\alpha_{3} \delta(1-\alpha_{1}-\alpha_{2}-\alpha_{3})\Phi_{3r}(\alpha_{1},\alpha_{2},\alpha_{3})e^{ipx(\alpha_{1}+v\alpha_{3})},
 \end {eqnarray}
 where $r$=$q,s$, $0\leq v\leq1 $ and $f_{3q}\approx f_{3s}\approx f_{3\pi}$ and
 \begin{equation}
 \fl \Phi_{3r}(\alpha_{i})=360\alpha_{1}\alpha_{2}\alpha_{3}^{2}\{1+\lambda_{3r} (\alpha_{1}-\alpha_{2})+\omega_{3r}\frac{1}{2}(7\alpha_{3}-3)\}.
 \end{equation}

\setcounter{section}{1}

\end{document}